\documentclass[11 pt]{article}
\usepackage{amssymb,amsfonts,amsmath}
\usepackage{graphicx}
\usepackage{dcolumn}
\usepackage{bm}
\usepackage{chemarr}

\begin{document}

\title{Investigating the robustness of the classical enzyme kinetic equations in small intracellular compartments}

\author{{\sc Ramon Grima} \\ \\ School of Biological Sciences \\
Centre for Systems Biology at Edinburgh, University of Edinburgh, UK}
\date{}
\maketitle \setlength{\baselineskip}{22pt}

\vspace{0.5cm} \hrulefill
\begin{itemize}
\item Corresponding author: Ramon Grima, email: ramon.grima@ed.ac.uk
\end{itemize}

\newpage

\noindent\hrulefill

\section*{Abstract} \textbf{Background}: Classical descriptions of enzyme
kinetics ignore the physical nature of the intracellular
environment. Main implicit assumptions behind such approaches are
that reactions occur in compartment volumes which are large enough
so that molecular discreteness can be ignored and that molecular
transport occurs via diffusion. Though these conditions are
frequently met in laboratory conditions, they are not characteristic
of the intracellular environment, which is compartmentalized at the
micron and submicron scales and in which active means of transport
play a significant role.
\\ \textbf{Results}: Starting from a master equation description of enzyme reaction kinetics
and assuming metabolic steady-state conditions, we derive novel mesoscopic rate
equations which take into account (i) the intrinsic molecular noise
due to the low copy number of molecules in intracellular
compartments (ii) the physical nature of the substrate transport
process, i.e. diffusion or vesicle-mediated transport. These
equations replace the conventional macroscopic and deterministic equations
in the context of intracellular kinetics. The latter are recovered in the limit of infinite
compartment volumes. We find that deviations from the predictions of classical kinetics are
pronounced (hundreds of percent in the estimate for the reaction
velocity) for enzyme reactions occurring in compartments which are
smaller than approximately 200nm, for the case of substrate
transport to the compartment being mediated principally by vesicle
or granule transport and in the presence of competitive enzyme
inhibitors.
\\ \textbf{Conclusions}: The derived mesoscopic rate equations describe
subcellular enzyme reaction kinetics, taking into account, for the first time,
the simultaneous influence of both intrinsic noise and the mode of transport.
They clearly show the range of applicability of the conventional deterministic equation models,
namely intracellular conditions compatible with diffusive transport and simple enzyme mechanisms in
several hundred nanometre-sized compartments. An active transport mechanism coupled with large intrinsic
noise in enzyme concentrations is shown to lead to huge deviations from the predictions of deterministic
models. This has implications for the common approach of modeling large intracellular reaction networks using
ordinary differential equations and also for the calculation of the effective dosage of competitive inhibitor drugs.
\newpage

\section*{Background}

The inside of a cell is a highly complex environment. In the past two decades, detailed measurements of the chemical and biophysical properties of the cytoplasm have established that the conditions in which intracellular reactions occur are, by and large, very different than those typically maintained in laboratory conditions. One of the outstanding differences between \emph{in vivo} and \emph{in vitro} conditions, is that in the former, biochemical reactions typically occur in minuscule reaction volumes \cite{LubyPhelps1}. For example, in eukaryotic cells, many biochemical pathways are sequestered within membrane-bound compartments, ranging from $\sim 50$nm diameter vesicles to the nucleus, which can be several microns in size \cite{Alberts}. It is also found that the total concentration of macromolecules inside both prokaryotic and eukaryotic cells is very large \cite{Minton,Trepat}, of the order of $50-400$ mg/ml which implies that between $5\%$ and $40\%$ of the total intracellular volume is physically occupied by these molecules \cite{SchnellTurner}. The concentration of these crowding molecules is highly heterogeneous (see for example \cite{Medalia}), meaning that typically one will find small pockets of intracellular space, characterized by low macromolecular crowding, surrounded by a ``sea'' of high crowding; such pockets of space may serve as effective compartments where reactions may occur more easily than in the rest of the cytosol. Analysis of experimental data for the dependence of diffusion coefficients with molecular size suggests the length scale of such effective compartments is in the range 35-50nm \cite{LubyPhelps2}, a size comparable to that of the smallest vesicles. The significant crowding also suggests that frequently an active means of transport such as vesicle-mediated transport, may be more desirable than simple diffusion as a means of intracellular transport.

The volume of a spherical cavity of space of diameter 50nm is merely $\sim6.5 \times 10^{-20}$ liters, an extremely small number compared to the typical macroscopic reaction volumes of \emph{in vitro} experiments (experimental attolitre biochemistry is still in its infancy - see for example \cite{Pick}). These very small reaction volumes imply that at physiologically relevant concentrations (nano to millimolar), the copy number of a significant number of intracellular molecules is very small \cite{LubyPhelps1} and consequently that intrinsic noise cannot be ignored; for example $255 \mu M$ corresponds to an average of just 10 molecules in a 50nm vesicle and fluctuations about this mean of the order of 3 molecules \cite{Grima1}.

The traditional mathematical framework of physical chemistry ignores the basic physical properties of the intracellular environment. Kinetics are described by a set of coupled ordinary differential equations which implicitly assume (i) that the reaction compartment is so large that molecular discreteness can be ignored and that hence integer numbers of molecules per unit volume can be replaced by a continuous variable, the molar concentration. Since the number of molecules is assumed to be very large, stochastic fluctuations are deemed negligible and the equations are hence deterministic; (ii) the reaction compartment is well-stirred so that homogeneous conditions prevail throughout \cite{Grima1}. Both assumptions can be justified for reactions occurring in a constantly stirred reactor of macroscopic dimensions. However if diffusion is the dominant transport process inside the compartment then the homogeneity assumption holds only if the volume is small enough so that in the time between successive reactions, a molecule will diffuse a distance much larger than the size of the compartment. This comes at the expense of the first assumption.
It hence appears natural that for intracellular applications, the first assumption, namely that of deterministic kinetics cannot be justified \emph{a priori}. The second assumption can be justified if reactions are localized in sufficiently small parts of the cell and in particular for reaction-limited processes i.e. those for which the typical time for two molecules to meet each other via diffusion is much less than the typical time for them to react if they are in close proximity. For such conditions, a molecule will come within reaction range several times before participating in a successful reaction, in the process sampling the compartment many times which naturally leads to well-mixed conditions \cite{Grima1,Gillespie,AvrahamHavlin}.

In this article we seek to understand the magnitude of deviations from the classical kinetic equations in small intracellular compartment volumes. We specifically focus on the case of reaction-limited enzyme reactions which allows us to relax the first assumption of physical chemistry while keeping the second one; this makes the mathematics tractable. We quantify deviations from classical kinetics in the context of the Michaelis-Menten (MM) equation; this is the cornerstone of present day enzyme kinetics and is a derivative of the traditional deterministic mathematical framework based on ordinary differential equations. In steady-state metabolic conditions, it is predicted to be exact. Thus this equation is ideal as a means to accurately test the effects of small-scale compartmentation on chemical kinetics. We consider three successive biological models of intracellular enzyme kinetics, each one building on the biological detail and realism from the previous one (Figure 1). The models incorporate the intrinsic noisiness of kinetics in small compartments, the details of the substrate transport process to the compartment (diffusion or active transport) and the presence of intra-compartmental molecules other than substrate molecules which may modulate the enzyme-catalyzed process e.g. inhibitors. On the macroscopic level, i.e. for large volumes, the steady-state kinetics of all models conform to the MM equation. We test whether this equation holds on the on the length scale of small intracellular compartments by deriving the dependence of the ensemble averaged rate of product formation on the ensemble-averaged substrate concentration from the corresponding stochastic models in the steady-state. It is shown via both calculation and stochastic simulation that at these small length scales the MM equation breaks down, being replaced by a new more general equation. Practical consequences of this breakdown are illustrated in the context of the calculation of the effective dosage of enzyme inhibitor drug needed to suppress intra-compartmental enzyme activity by a given amount. To make our approach accessible to readers not familiar with stochastic equations and their analysis, in the Results/Discussion sections we mainly focus on the biological/biophysical context and implications of the models together with the main mathematical results which are verified by simulation. Detailed mathematical derivations and the methods of simulation are relegated to the Methods section.

\section*{Results}

\subsection*{Model I: Michaelis-Menten reaction occurring in a compartment volume of sub-micron dimensions. Substrate input into compartment occurs via a Poisson process} This is the simplest, biologically-relevant case (Figure 1(A)). The reaction scheme is $\xrightarrow{k_{in}} S + E  \xrightleftharpoons[k_{1}]{k_0}  \ C \xrightarrow{k_2} E + P$. Substrate molecules (\emph{S}) are continuously supplied inside the compartment at some rate $k_{in}$, they reversibly bind to enzyme molecules (\emph{E}) with rate constants $k_0$ (forward reaction) and $k_{1}$ (backward reaction) to form transitory enzyme-substrate complex molecules (\emph{C}) which then decay with rate $k_2$ into enzyme and product molecules (\emph{P}). The substrate input is assumed to be governed by a Poisson process with mean $k_{in}$; this is consistent with substrate transport to the compartment being dominated by normal diffusion. The enzyme acts as a catalyst, effectively speeding up the reaction by orders of magnitude. It is assumed that diffusion inside the compartment is normal and not rate-limiting on the catalytic process i.e. well-mixed conditions or rate-limited kinetics inside the compartment. Given these conditions we ask ourselves what is the relationship between the reaction velocity and the substrate concentration inside the compartment. The simplest approach consists of writing down the rate equations of traditional physical chemistry:
\begin{align}
[E_T] & = [E]+[C]=constant, \\
{d[S]}/{dt} & = k_{in} - k_0 [E][S] + k_1 [C], \\
{d[C]}/{dt} & = k_0 [E][S] - (k_1+k_2)[C], \\
\frac{d[P]}{dt} & = k_2 [C].
\end{align}
By imposing steady-state conditions we get the sought-after relationship which is simply the well-known MM equation:
\begin{equation}
\frac{d[P]}{dt} = v = \frac{v_{max}[S]}{K_M+[S]},
\end{equation}
where $K_M=(k_1+k_2)/k_0$ is the MM constant, $v_{max} = k_2 [E_T]$ is the maximum reaction velocity and square brackets indicate the macroscopic concentrations. We note that steady-state conditions for substrate necessarily require that $k_{in} \le v_{max}$ otherwise the substrate will continuously accumulate with time. Though this approach is simple and straightforward, as mentioned in the introduction, the assumptions behind the formulation of the rate equations are not consistent with the known physical properties of the cytoplasm. In particular it is clear that if the volume of our compartment is very small (as is the case), the numbers of particles is also quite small, meaning that the concept of a continuous variable such as the average macroscopic concentration has little meaning. Rather we require a mathematical description in terms of discrete, integer numbers of particles and which is stochastic. The natural description of such cases is a master equation which is a differential equation in the joint probability function $\pi$ describing the system \cite{VanKampen,Bartholomay1,Bartholomay2,Jachimowski}:
\begin{align}
\nonumber
\frac{d\pi}{dt} & = k_{in} \Omega (\Theta_S^{-1}-1)\pi + \frac{k_0}{\Omega} (\Theta_S \Theta_C^{-1}-1) n_S n_E \pi\ \\
&+ k_1 (\Theta_C \Theta_S^{-1}-1) n_C \pi + k_2 (\Theta_C \Theta_P^{-1}-1) n_C \pi,
\end{align}
where $\pi=\pi(n_S,n_C,n_P)$, $n_Y$ is the integer number of molecules of type $Y$, $\Omega$ is the compartment volume, and $\Theta_X^{\pm 1}$ are step operators defined in the Methods section. This equation cannot be solved exactly. However it can be solved perturbatively using the system-size expansion due to van Kampen \cite{VanKampen}. This expansion is one in powers of the inverse square root of the compartment volume. In the Methods section, we calculate the first three terms of the expansion, namely those proportional to $\Omega^{1/2}$, $\Omega^0$ and $\Omega^{-1/2}$. The first term, being the dominant one for large volumes, gives back as expected, the rate equations Eqs. (1)-(4). The second term gives the magnitude of stochastic fluctuations about the macroscopic concentrations. Corrections to the rate equations and to the MM equation (due to small compartment volumes or equivalently due to intrinsic noise) are found by considering the third term. In the rest of the article, instead of using the reaction velocity $v$, we use the normalized reaction velocity, $\alpha$, which is simply the velocity of the reaction, $v$, divided by the maximum reaction velocity, $v_{max}$. Given some measured intracompartmental substrate concentration, $[S^*]=\langle n_S/\Omega \rangle$ (angled brackets imply average), the relationship between the normalized reaction velocity predicted by the MM equation ($\alpha_M=[S^*]/(K_M+[S^*])$) and the actual normalized reaction velocity ($\alpha$), as predicted by our theory, is given by:
\begin{equation}
\alpha + (1-\alpha_M)f(\alpha) \Omega^{-1} = \alpha_M,
\end{equation}
where,
\begin{equation}
f(\alpha) = \frac{\alpha^2}{K_M+[E_T](1-\alpha)^2}.
\end{equation}
Hence the prediction of the MM equation is only correct, i.e. $\alpha = \alpha_M$, in the limit of infinitely large compartment volumes, in which case the second term on the left hand side of Eq. (7) will become vanishingly small and can be neglected. For finite compartment volumes, the MM equation is not exact (except in the two limiting cases of $\alpha_M \rightarrow 0$ and $\alpha_M \rightarrow 1$) but is at best an approximation, even though steady-state conditions are imposed; this is at odds with the prediction of the conventional deterministic theory. An inspection of Eqs. (7) and (8) shows that the magnitude of the deviations from the MM equation depends on the two non-dimensional quantities: (i) $K_M \Omega$, a measure of the rate at which enzyme-substrate combination events occur relative to the rate of decay of complex molecules; (ii) $[E_T] \Omega$, the total integer number of enzyme molecules in the compartment.

As shown in the Methods section, the MM equation is found to implicity
assume that the noise about the macroscopic substrate and enzyme
concentrations is uncorrelated (this assumption has generally been found to
be at the heart of many macroscopic models - for example see \cite{Grima2}); properly taking into account these
non-zero correlations leads to the corrections encapsulated by
Eqs. (7) and (8). These correlations are expected to be small
in two particular cases: (i) if $K_M$ is large; in this case when
substrate molecules combine with an enzyme to form a complex, the
latter dissociates very quickly back into free enzyme and thus
successive enzyme-substrate events to the same enzyme molecule are
bound to be almost independent of each other. The opposite situation
of small $K_M$ would imply that the bottleneck in the catalytic
process is the decay of complex rather than enzyme-substrate
combination; if a successful combination occurs, the next substrate
to arrive to the same enzyme molecule would have to wait until the
complex decays, naturally leading to correlations between successive
enzyme-substrate combination events. (ii) if the total number of
enzyme molecules is large; in such a case, at any one time, the
noise about the macroscopic concentrations will be the sum total
from a large number of enzymes, each at a different stage in the
catalytic process and each independent from all others, which
naturally dilutes any temporal correlations.

To estimate the magnitude of the deviations from the MM equation inside cells, we use the above two equations, Eqs. (7) and (8), to compute the absolute percentage error $R_p = 100|1 - \alpha_M/\alpha|$. These estimates are also computed by stochastic simulation of the master Eq. (6), using the exact stochastic simulation algorithm of Gillespie \cite{Gillespie} (see Methods for details regarding the method of simulation); this provides a direct test of the theory. Figure 2 shows the typical dependence of $R_p$ on $\alpha_M$, as predicted by both theory (solid lines) and simulation (data points). Generally the agreement between the two is found to be very good; discrepancies increase as $K_M$ and compartment volume decrease but are small for parameter values realistic for intracellular conditions. The maxima of such plots gives the maximum absolute percentage error which is a measure of the maximum expected deviations from the MM equation. Table 1 summarizes these estimates (theory and simulation) over wide ranges of the parameters typical of \emph{in vivo} conditions: $K_M=10\mu M-1000\mu M$ \cite{Berg}, enzyme copy numbers of ten and one hundred per compartment which correspond to enzyme concentrations ranging from $4\mu M$ to 2.5\emph{m}M and compartment diameters ranging from 50nm to 200nm. Note that the maximum deviations from the MM equation are estimated to be less than approximately $20\%$ and typically just a few percent over large ranges of parameter values -- this robustness of the MM equation with respect to intrinsic molecular noise is indeed surprising, since strictly speaking it is only valid for infinite compartment volumes.

The theory is always found to underestimate the actual deviations predicted by simulations; hence the theoretical expressions provide a quick, convenient way by which one can generally estimate a lower bound on the deviations to be expected from the MM equation without the need to perform extensive stochastic simulation.

\subsection*{Model II: Michaelis-Menten reaction occurring in a compartment volume of sub-micron dimensions. Substrate is input into compartment in groups or bursts of M molecules at a time}

Model I captures the basics of a general enzyme-catalyzed process occurring in a small intracellular compartment. In this section we build upon this model to incorporate further biological realism.  In particular, in the previous model we assumed that substrate input can be well described by a Poisson process, where one molecule at a time is fed into the compartment with some average rate $k_{in}$. This is the simplest possible assumption and approximates well the situation in which molecules are brought to the compartment via normal diffusion. However there are many situations where this may not be the case; we now describe two such cases.

The intracellular condition of macromolecular crowding limits the Brownian motion of molecules in the cytoplasm, this being reflected in the relatively small diffusion coefficients measured \emph{in vivo} compared to those known \emph{in vitro} for moderately to relatively large molecules. Experiments with inert tracer particles in the cytoplasm of Swiss 3T3 cells show that the \emph{in vivo} diffusion coefficient is an order of magnitude less than that \emph{in vitro} for molecules with hydrodynamic radius $14$nm and diffusion becomes negligibly small for molecules larger than approximately $25$nm \cite{LubyPhelps2}; similar results have been obtained in Xenopus neurons \cite{Popov} and skeletal muscle myotubes \cite{ArrioDupont}. If diffusion is considerably hindered, one expects active transport to become a more desirable mode of transport. Indeed there exists ample evidence for the active transport of macromolecules: they are typically packaged in a vesicle or a granule which is then transported along microtubules or by some other means. It is also found that each vesicle or granule typically contains several of these molecules (examples are: mRNA molecules - several estimated per granule \cite{Ainger,BassellSinger}; cholesterol molecules which are transported in low-density lipoproteins \cite{Alberts} - approximately 1500 per lipoprotein).

Generally an active means of transport is not exclusively linked with the transport of large substrate molecules. The cell being a highly compartmentalized and dynamic entity requires for its survival the precise transport of certain molecules from one compartment to another and a regulation of this transport depending on its current physiological state. Brownian motion leads to an isotropic movement of molecules down the concentration gradient and to a consequent damping of the substrate concentration with distance. In contrast active transport provides a directed (anisotropic) means of transport with little or no loss of substrate with distance, is independent of the concentration gradient and it is also easily amenable to modulation.

Hence it follows that a more general process modeling molecular entry into an intracellular compartment is one in which $M$ molecules are fed into the compartment at a rate $k_{in}^0$; the latter rate constant is the rate at which vesicles or granules arrive to the site of the compartment (Figure 1(B)). The total mean substrate input rate is then $k_{in}=Mk_{in}^0$. The special case $M=1$ corresponds to Model I. We construct the relevant master equation and employ the system-size expansion as for the previous model (see Methods for details); it is found that the deterministic rate equations are exactly Eqs. (1)-(4) i.e. at the macroscopic level, given two reactions occurring in two different compartments, both with the same total mean substrate input rate $k_{in}$ but one occurring via diffusion (e.g. $M = 1, k_{in}^0 = 1$) and the other via active transport (e.g. $M = 10, k_{in}^0 = 0.1$) , cannot be distinguished. However if the compartment volumes become small, then once again we find corrections to the MM equation and interestingly these corrections are sensitive to the mode of transport. The relationship between the normalized reaction velocity predicted by the MM equation ($\alpha_M$) and the actual normalized reaction velocity ($\alpha$), as predicted by our theory, is given by Eq. (7) together with:
\begin{equation}
f(\alpha) = \frac{ \alpha [\alpha
+\frac{1}{2}(M-1)]}{K_m+[E_T](1-\alpha)^2}.
\end{equation}
This suggests that generally deviations from the predictions of
the MM equation increase with the carrying capacity, $M$, of the
vesicle or granule. To compare the effects of active transport and
diffusion on the kinetics, we set $M=50$ and adjusted $k_{in}^0$ so
that in all cases, the total mean substrate input rate for model II
is equal to $k_{in}$, the input rate of Model I (i.e. the
two models would be indistinguishable from a macroscopic point of
view). Using the same procedure as for Model I, we computed the maximum
percentage error using Eqs. (7) and (9) and also from simulations. The
results are summarized in Table 2. Notice that now the deviations
from the MM equation are much larger than before, running into
hundreds of percent rather than the tens as for Model I. Because of
the increase in substrate fluctuations, the quantitative accuracy of
the theory is now less than before; it generally fares very well for
compartments with diameters larger than $\sim100$nm and $K_M$ larger
than $\sim100\mu M$. Nevertheless in all cases theory does correctly
predict a large increase in discrepancy between the reaction
velocities given by the deterministic MM equation and those from
stochastic simulation compared to the case of Model I. The intuitive
reason behind these increases in discrepancy is that substrate which
is input in bursts enhances correlations between successive enzyme-substrate events.

The explicit dependence of the reaction velocity on substrate concentration is complex and generally requires the solution of the cubic polynomial encapsulated by Eqs. (7) and (9). However for small substrate concentrations, the equations simplify to a simple linear equation:
\begin{equation}
\alpha = [S^*]\left(K_M \left[1+\frac{M-1}{2 \Omega (K_M+[E_T])} \right]\right)^{-1}
\end{equation}
Note that if the MM equation was correct, one would expect $\alpha =
[S^*]/K_M$. Indeed Eq. (10) reduces to the latter prediction in the
limit of large volumes. Note also that this renormalization of the
proportionality constant occurs only if the substrate input occurs
in bursts, i.e. $M > 1$. These predictions of our theory are
verified by simulations (Figure 3).

\subsection*{Model III: Michaelis-Menten reaction with competitive inhibitor occurring in a compartment volume of sub-micron dimensions. Substrate input as in previous models}

In this last section, we further build on the previous two models by adding competitive inhibitors to the intracellular compartment in which enzymes are localized. A competitive inhibitor, $I$, is one which binds reversibly to the active site of the enzyme (forming a complex $EI$), thus preventing substrate molecules from binding to the enzyme and slowing down catalysis (Figure 1(C)). In standard textbooks and in the literature, this is typically modeled by the set of reactions (see for example \cite{Fersht}): $\xrightarrow{k_{in}} S + E \xrightleftharpoons[k_{1}]{k_0}  \ C
\xrightarrow{k_2} E + P, \  E\xrightleftharpoons[k_3]{k_4} EI$, where $k_4 = k_4^0 [I]$ and $[I]$ is the inhibitor concentration. Note that it is implicitly assumed that inhibitor is in such abundance that the second-order bimolecular reaction between inhibitor and enzyme can be replaced by a pseudo first-order reaction with constant inhibitor concentration. We shall assume the same in our model. Substrate input into the compartment is considered to occur as in Model II since this encapsulates that of Model I as well. The deterministic model of this set of reactions leads to a MM equation of the form:
\begin{equation}
\frac{d[P]}{dt} = \frac{v_{max}[S]}{K_m(1+\beta)+[S]},
\end{equation}
where $\beta = [I]/K_i$ and $K_i=k_3/k_4^0$ is the dissociation
constant of the inhibitor. The perturbative solution of the master
equation describing the system is now significantly more involved
than in previous models; the underlying reason for this is that the
computation of the noise correlators to order $\Omega^0$ requires
the inversion of a $6 \times 6$ matrix as opposed to a $3 \times 3$ one in
previous models (see Methods for details). The analysis predicts
corrections to the MM equation by postulating a new mesoscopic rate
equation having the form of Eq. (7) together with:
\begin{equation}
f(\alpha) = \frac{1+\beta}{K_m [E_T]} \frac{\sum_{i=0}^4 c_i
(1-\alpha)^i}{\sum_{i=0}^4 d_i (1-\alpha)^i},
\end{equation}
where $c_i$ and $d_i$ are coefficients with a complex dependence on
the various enzyme parameters (these are given in full in the Methods Section).
Table 3 shows the maximum percentage error
computed using Eqs. (7) and (12) and also from
simulations for the cases in which substrate input occur a molecule
at a time and in bursts of 50 at a time. The parameter values chosen
in the simulations and calculations (see caption of Table 3) are
typical for many enzymatic processes: the bimolecular rate
coefficients span the range $10^6-10^9 s^{-1} M^{-1}$ \cite{Fersht},
the backward decay processes are in the middle of the range $10 -
10^5 s^{-1}$ \cite{Fersht}, the inhibitor concentration is ten times
larger than the total enzyme concentration (satisfying the implicit assumption that
the inhibitor is in significantly larger concentration than free enzyme), and the
intracompartmental enzyme concentrations are in the range $4-255\mu
M$. The deviations from the MM equation in this case are more severe
than in the previous two models, this being due to non-zero
correlations between substrate and the complex $EI$ in addition to
the already present correlations between substrate and complex $C$.
Note that the agreement between theory and simulations is overall
better than in previous models, even when the burst size is large,
$M = 50$. As mentioned in the section for Model I, discrepancies between theory
and simulation are generally found to decrease with increasing $K_M$; for the case of competitive inhibition,
the effective $K_M$ of the reaction is significantly larger than
that of the enzyme (see Eq. (11)), which explains the increased agreement between theory and
simulations for Model III compared to the previous two models.

A significant number of drugs suppress a chain of biochemical
reactions by reducing the activity of key enzymes in the pathway via
competitive inhibition \cite{Berg}. The conventional method to estimate
the required concentrations of these inhibitors involves
plotting the variation of the enzyme activity with inhibitor
concentration, $[I]$, using the MM equation; the substrate
concentration is kept fixed and is chosen so that at $[I]=0$, the
reaction velocity is close to the maximum, $v_{max}$. Since there
are significant corrections to the MM equation when reactions occur
in intracellular compartments, it is not clear how accurate are
estimates of $[I]$ based upon it. Figure 4 compares the enzyme
activity curve based on the MM equation with the theoretical
predictions for the corrected enzyme activity curves based on the
mesoscopic rate equation embodied by Eqs. (7) and (12), for
compartments of diameter 50nm and 100nm (inset) and for substrate
input burst sizes of $M=20$ and $50$. The substrate concentration is
chosen so that at $[I] = 0$, $v/v_{max}=0.909$ in all cases. We find
that generally as the burst size increases, the actual inhibitor
concentration needed to suppress enzyme activity by a given amount
is larger than that estimated by the MM equation; this discrepancy
decreases with increasing compartment volume. For the example in
Figure 4, for the case in which substrate is input into the
compartment in bursts of $M=50$, the actual inhibitor concentration
needed to decrease the enzyme activity from $0.909$ to $0.1$ is
approximately 7 times larger than the MM estimate; if the
compartment diameter is doubled (inset of Figure 4), the actual
inhibitor concentration needed is less than twice that of the MM
estimate. Generally we find that for the typical parameter values of
enzymatic reactions, the corrections to the enzyme-activity curves
can be neglected for compartments larger than about $200$nm in
diameter.

\section*{Discussion and Conclusion}

In this last section we discuss some fine points regarding: (i)the
assumptions behind the use of master equations which throws light on
the range of use of the derived mesoscopic equations, (ii) the use
of the system-size expansion to perturbatively solve the master
equation and (iii) the assumption of steady-state metabolic
conditions. We conclude by placing our work in the context of
previous recent studies of stochastic enzyme kinetics and discuss
possible experiments to verify some of the conclusions we have
reached.

We have implicitly assumed throughout the article that a single
(global) master equation model suffices to capture the deviations
from classical kinetics due to fluctuations in chemical
concentrations inside a single subcellular compartment. As noted by
Baras and Mansour \cite{Baras}, ``the global master equation selects
the very limited class of exceptionally large fluctuations that
appear at the level of the entire system, disregarding important
nonequilibrium features originated by local fluctuations.'' Hence
the results presented here necessarily underestimate the possible
deviations from classical kinetics, in particular the local
fluctuations due to diffusion of molecules inside the compartment.
These local fluctuations are typically small for reaction-limited
processes (as in this article) but significant for diffusion-limited
ones. To capture them effectively, one would be required to
spatially discretize the compartment into many small elements and
describe the reaction-diffusion processes between these elements by
means of a multivariate master equation \cite{VanKampen,Baras}. The
latter is known as a reaction-diffusion master equation; typically
it does not allow detailed analytical investigation as for a global
master equation and one is limited to stochastic simulation. Use of
the global master equation is also restricted for compartments which
are not too small: in particular the linear dimensions of the
compartment should be larger than the average distance traveled by a
molecule before undergoing a successful reaction with another
molecule i.e. the length scale is much larger than that inherent in
molecular dynamics simulation \cite{Baras}.

We have applied the systematic expansion due to van Kampen to
perturbatively solve the master equation. It is sometimes \emph{a
priori} assumed that because this expansion is about the macroscopic
concentrations, it cannot give information regarding the stochastic
kinetics of few particle / small volume systems. This is true if one
restricts oneself to the expansion to order $\Omega^{0}$ i.e. the
linear-noise approximation; this is commonly the case found in the
literature since the algebra becomes tedious if one considers more
terms. However we note that as argued and shown by van Kampen
himself \cite{VanKampen}, terms beyond the linear-noise
approximation in the system-size expansion add terms to the
fluctuations that are of order of a single particle relative to the
macroscopic quantities and are essential to understanding how
fluctuations are affected by the presence of non-linear terms in the
macroscopic equation (substrate-enzyme binding in our case). In our
theory we went beyond the linear-noise approximation. We find that
the predicted theoretical results are in reasonable agreement, in
many cases (comparison of bold and italic values in Tables 1, 2 and 3),
with stochastic simulations of just a few tens of enzyme
molecules in sub-micron compartments, which justifies our methodology.

We have also imposed metabolic steady-state conditions inside the
subcellular compartment. Technically this is convenient since in
such a case one does not deal with complex transients. Also since
under such conditions the MM equation is exact from a deterministic
point of view, it provides a very useful reference point versus
which to accurately compute deviations due to intrinsic noise. In
reality one may not always have steady-state conditions inside
cells, this depending strongly on the rate of substrate input
relative to the maximum rate at which the enzyme can process
substrate. Another possibility is that one is dealing with a batch
reaction i.e. one in which a number of substrate molecules are
transported at one go and just once to the subcellular compartment
(e.g. via vesicle-mediated transport) and the reaction proceeds
thereafter without any further substrate replenishment. This latter
scenario is compatible with the presentation of the MM equation
typical in standard physical chemistry textbooks. The MM equation is
then an approximation (not exact as in steady-state case) to the
deterministic kinetics, when substrate is present in much larger
concentration than enzyme. This case is currently under
investigation using the same perturbative framework used in this
article.

We note that this is not the first attempt to study stochastic
enzyme kinetics. The bulk of recent studies
\cite{English,Kou,Stefanini,Qian} have focused on understanding the
kinetics of a Michaelis-Menten type reaction catalyzed by a single
enzyme molecule. Deviations from classical kinetics were found to be
most pronounced when one takes into account substrate fluctuations
\cite{Stefanini}. These pioneering studies were restricted to a
single-enzyme assisted reaction which reduces complexity thereby
making it ideal from a theoretical perspective; since the reaction
is dependent on just a single enzyme molecule one also finds maximum
deviations from deterministic kinetics. In reality, it is unlikely
to find just one enzyme molecule inside a subcellular compartment -
as mentioned in the introduction a physiological concentration of
just a few hundred micromolar would correspond to few tens inside
the typically smallest subcellular compartment. It is also the case
that diffusion may not always be the main means of substrate
transport to the compartment and that the reaction maybe more
complex than the simple Michaelis-Menten type reaction of these
previous studies. The present study fills in these gaps by using a
systematic method to derive approximate and relatively simple
analytic expressions for mesoscopic rate equations describing the
kinetics of the general case of $N$ enzyme molecules in a
subcellular compartment with or without active transport of
substrate and in the presence of enzyme inhibitors. Most importantly
our approach shows the effects of intrinsic noise on the kinetics
can be captured via effective ordinary differential equations. This
enables quick estimation of the magnitude of stochastic effects on
reaction kinetics and thus gives insight into whether a model or
parts of a model should be designed to be stochastic or
deterministic without the need for extensive stochastic simulation.
In the present study, this approach enabled us to readily compute,
for the first time, the deviations from deterministic kinetics for a
broad range of realistic \emph{in vivo} parameter constants (Tables 1, 
2 and 3), a task which would be considerably lengthy if one had to
rely solely on data obtained from ensemble-averaged stochastic
simulations.

We conclude by briefly discussing possible experiments which can
verify the predictions made in this article. It is arguably not an
easy task to perform the required experiments in real-time in a
living cell. A viable alternative would consist of monitoring
reaction kinetics inside single artificially-made vesicles. Pick et
al \cite{Pick} have shown that the addition of cytochalasin to
mammalian cells induces them to extrude from their plasma membrane
minuscule vesicles of attolitre volume with fully functional cell
surface receptors and also retaining cytosolic proteins in their
interior. The change in the intra-vesicular calcium ion concentration in response
to surface ligand binding was measured using fluorescence confocal
microscopy (FCM). Since the vesicle sizes are of typical small
sub-cellular compartment dimensions (1 attolitre corresponds to a
spherical vesicle of approximate diameter 120nm) and FCM allows the
measurement of the concentration of a fluorescent probe (via a
calibration procedure), this experimental technique appears ideal to
verify the predictions of Model I and of Model III for the case of
diffusive substrate transport. Model II and Model III with
vesicle-transport of substrate are probably much more challenging to
verify since one then needs to construct the \emph{in vitro}
equivalent of microtubules. This is within the scope of synthetic
biology and may be a possibility in the next few years.

\section*{Methods}

We here provide full details of the calculations reported in the
Results section. The system size-expansion which is at the heart of the
analysis has to-date not been applied extensively to biological problems
and thus we go into some detail in its elucidation in Subsection I,
which is dedicated exclusively to Model I. For other recent applications of
the general method in the context of reaction kinetics, see for example \cite{Grima3} and \cite{McKane}.
Subsections II and III (treating Model II and Model III, respectively) naturally build
on the results of the first subsection and thus we only give the main
steps of the calculations in these last two cases. Subsection IV has a brief
discussion of the simulation methods used to verify the theoretical
results.

\subsection*{Model I: Michaelis-Menten reaction occurring in a compartment volume
of sub-micron dimensions. Substrate input into compartment is
modeled as a Poisson process}

The reaction scheme is $\xrightarrow{k_{in}} S + E
\xrightleftharpoons[k_{1}]{k_0}  \ C \xrightarrow{k_2} E + P$.
The stochastic description of this system is encapsulated by the master
equation which is a differential equation in the joint probability
function $\pi$ describing the system:
\begin{align}
\frac{d\pi}{dt} & = k_{in} \Omega (\Theta_S^{-1}-1)\pi +
\frac{k_0}{\Omega} (\Theta_S \Theta_C^{-1}-1) n_S n_E \pi\\ \nonumber &+ k_1
(\Theta_C \Theta_S^{-1}-1) n_C \pi + k_2 (\Theta_C \Theta_P^{-1}-1) n_C \pi,
\end{align}
where $\pi=\pi(n_C,n_P,n_S)$, $n_X$ is the integer number of
molecules of type $X$ (where $X={C,P,S}$), $\Omega$ is the
compartment volume, and $\Theta_X^{\pm 1}$ are the step operators defined
by their action on a general function $g(n_X)$ as: $\Theta_X^{\pm 1}
g(n_X) = g(n_X \pm 1)$. Note that the relevant variables are three,
not four: the integer number of molecules of free enzyme ($n_E$) is
not an independent variable due to the fact that the total amount of
enzyme is conserved. The master equation cannot be solved exactly
but it is possible to systematically
approximate it by using an expansion in powers of the inverse square
root of the volume of the compartments. This is
generally called the system-size expansion \cite{VanKampen}.

The method proceeds as follows. The stochastic quantity,
$n_X/\Omega$, fluctuates about the macroscopic concentrations [X];
these fluctuations are of the order of the square root of the number
of particles:
\begin{equation}
n_X = \Omega [X] + \Omega^{1/2} \epsilon_X.
\end{equation}
Note that since $n_E + n_C = constant$, it follows that $n_E =
\Omega [E] - \Omega^{1/2} \epsilon_C$. The joint distribution
function and the operators can now be written as functions of the
new variables, $\epsilon_X$, giving: $\pi =
\Pi(\epsilon_C,\epsilon_P,\epsilon_S,t)$ and $\Theta_X^{\pm 1} = 1 \pm \Omega^{-1/2}
{\partial}/{\partial\epsilon_X}+\frac{1}{2}\Omega^{-1}
{\partial^2}/{\partial\epsilon_X^2} + O(\Omega^{-3/2}) $; using
these new variables the master equation Eq. (13) takes the form:
\begin{align}
\frac{\partial \Pi}{\partial t} - \Omega^{1/2}
\biggl(\frac{d[C]}{dt} \frac{\partial \Pi}{\partial \epsilon_C} +
\frac{d[P]}{dt} \frac{\partial \Pi}{\partial \epsilon_P} +
\frac{d[S]}{dt} \frac{\partial \Pi}{\partial \epsilon_S} \biggr) =
\Omega^{1/2} a_1 \Pi + \Omega^{0} a_2 \Pi + \Omega^{-1/2} a_3 \Pi +
O(\Omega^{-1})
\end{align}
where
\begin{align}
a_1 = -(k_{in} + k_1 [C] - k_0 [E][S]) \frac{\partial}{\partial
\epsilon_S} + ((k_1+k_2)[C] - k_0 [E][S])\frac{\partial }{\partial
\epsilon_C} - k_2[C] \frac{\partial}{\partial \epsilon_P},
\end{align}

\begin{align}
\nonumber a_2 = &\frac{1}{2}k_{in}
\frac{\partial^2}{\partial\epsilon_S^2} + \frac{1}{2}
\biggl(\frac{\partial}{\partial \epsilon_S}-\frac{\partial}{\partial
\epsilon_C}\biggr)^2 (k_0 [S][E]+k_1 [C]) + k_2 \biggl[\frac{\partial}{\partial \epsilon_C} - \frac{\partial}{\partial \epsilon_P} \biggr] \epsilon_C
\\ &+ \biggl[\frac{\partial}{\partial \epsilon_S} -
\frac{\partial}{\partial \epsilon_C} \biggr] [k_0(\epsilon_S
[E]-\epsilon_C [S])- k_1 \epsilon_C]  + \frac{1}{2} k_2 \biggl(\frac{\partial}{\partial \epsilon_P}-\frac{\partial}{\partial \epsilon_C}\biggr)^2 [C],
\end{align}

\begin{align}
\nonumber a_3 = &\frac{1}{2} \biggl(\frac{\partial}{\partial
\epsilon_S}-\frac{\partial}{\partial \epsilon_C}\biggr)^2(k_0
\epsilon_S [E] - k_0 \epsilon_C [S] + k_1 \epsilon_C) \nonumber \\&- k_0
\biggl[\frac{\partial}{\partial \epsilon_S}
-\frac{\partial}{\partial \epsilon_C} \biggr] \epsilon_S \epsilon_C
+ \frac{1}{2} k_2 \biggl(\frac{\partial}{\partial
\epsilon_P}-\frac{\partial}{\partial \epsilon_C}\biggr)^2 \epsilon_C.
\end{align}

Note that in Eq. (18) terms which involve products of first and second-order
derivatives, third-order derivatives or higher have been omitted -
these do not affect the low-order moment equations which we will be
calculating.

\subsubsection*{Analysis of $\Omega^{1/2}$ terms}

The terms of order $\Omega^{1/2}$ are the dominant ones in the limit
of large volumes. By equating both terms of this order on the right
and left hand sides of Eq. (15) and using Eq. (16), one gets the
deterministic rate equations:
\begin{align}
{d[S]}/{dt} & = k_{in} - k_0 [E][S] + k_1 [C], \\
{d[C]}/{dt} & = k_0 [E][S] - (k_1+k_2)[C], \\
{d[P]}/{dt} & = k_2 [C].
\end{align}
These are exactly those which one would write down based on the
classical approach whereby one ignores molecular discreteness and
fluctuations. This is an important benchmark of the method since it
shows that it gives the correct result in the limit of large
volumes. On a more technical note, the cancelation of these two
terms of order $\Omega^{1/2}$ makes Eq. (15) a proper expansion in
powers of $\Omega^{-1/2}$. By imposing steady-state conditions we
have the Michaelis-Menten (MM) equation:
\begin{equation}
\frac{d[P]}{dt} = \frac{v_{max}[S]}{K_M+[S]},
\end{equation}
where $v_{max} = k_2 [E_T]$ is the maximum reaction velocity, $[E_T]
= [E]+[C]$ is the total enzyme concentration which is a constant at
all times and $K_M=(k_1+k_2)/k_0$ is the Michaelis-Menten constant.

\subsubsection*{Analysis of $\Omega^{0}$ terms}

To this order, the master equation is a multivariate
Fokker-Planck equation whose solution is Gaussian and thus fully
determined by its first and second moments. The equations of motion
for these moments can be straightforwardly obtained from the master
equation to this order, leading to a set of coupled but solvable
ordinary differential equations:

\begin{equation}
\partial_t \left[\begin{array}{c} \langle \epsilon_S \rangle \\ \langle \epsilon_C \rangle
\end{array}\right]=\left(\begin{array}{cc} -k_0 [E] & k_1 + k_0 [S]\\ k_0
[E] & -k_0(K_M+ [S])
\end{array}\right) \left[\begin{array}{c} \langle \epsilon_S \rangle \\ \langle \epsilon_C \rangle
\end{array}\right]
\end{equation}

\begin{equation}
\partial_t \left[\begin{array}{c} \langle \epsilon_S^2 \rangle \\ \langle
\epsilon_C^2 \rangle \\ \langle \epsilon_{S} \epsilon_{C} \rangle
\end{array}\right]=A \cdot \left[\begin{array}{c} \langle \epsilon_S^2 \rangle \\ \langle
\epsilon_C^2 \rangle \\ \langle \epsilon_{S} \epsilon_{C} \rangle
\end{array}\right] + B,
\end{equation}
where,
\begin{equation}
A = \left(\begin{array}{ccc} -2k_0 [E] & 0 & 2(k_1 + k_0 [S])\\ 0 &
-2k_0(K_M+ [S]) & 2k_0 [E] \\ 0 & -2k_2 & -2k_2 \end{array}\right) , B = \left(\begin{array}{c} k_{in}+k_1[C]+k_0[S][E] \\ k_0 ([S][E] + K_M[C]) \\
k_{in} + k_2 [C] \end{array}\right).
\end{equation}
Note that the matrices and vectors in the above equations have been
reduced to a simpler form by the application of a few row
operations. Note also that these equations are independent of
$\epsilon_p$ since the product-forming step is irreversible and
hence the fluctuations in substrate and complex are necessarily
decoupled from its fluctuations. At the steady-state, it is found that
$\langle \epsilon_{S,C} \rangle \rightarrow 0$. From Eq. (14), it is
clear that this implies that to this order the average number of substrate
molecules per unit volume, $\langle n_S/\Omega \rangle$, is simply
equal to the macroscopic concentration, $[S]$. The same applies for
complex molecules. Hence to this order in the system-size expansion
there cannot be any corrections to the macroscopic equations or to
the MM equation. By writing the macroscopic concentrations in
Eqs. (24) and (25) in terms of $k_{in}$ and solving, one
obtains the variance and covariance of the fluctuations about the
steady-state macroscopic concentrations. We here only give the
result for the covariance since this will be central to our
discussion later on:
\begin{equation}
\langle \epsilon_C \epsilon_S \rangle =
\frac{K_M[E_T]\alpha^2}{K_M+[E_T](1-\alpha)^2},
\end{equation}
where $\alpha = k_{in} / v_{max}$ is the normalized reaction
velocity of the enzyme.

\subsubsection*{Analysis of $\Omega^{-1/2}$ terms}

The system-size expansion is almost never carried out to this order
because of the algebraic complexity typically involved, however it
is crucial to find finite volume corrections to the deterministic
rate equations and in particular to the MM equation. Using the
master equation to this order, the first moment of the complex
concentration is governed by the equation of motion:
\begin{align}
{d \langle \epsilon_C \rangle}/{dt}  = -k_0([S] + K_M) \langle
\epsilon_C\rangle + k_0 [E] \langle \epsilon_S \rangle - k_0
\Omega^{-1/2} \langle \epsilon_S \epsilon_C \rangle.
\end{align}
Now the production of product \emph{P} from complex occurs through a decay
process which necessarily has to be described by a linear term of
the form: $k_{in}=k_2 \langle n_C/\Omega \rangle$ (the steady-state condition). Since the
steady-state macroscopic complex concentration is equal to
$[C]=k_{in}/k_2$, then it follows that to any order in the expansion
we have $\langle \epsilon_C \rangle = 0$. This is always found to be
the case in simulations as well. Hence it immediately follows from
Eq. (27) that the average of fluctuations about the macroscopic substrate
concentration are non-zero and given by:
\begin{equation}
\langle \epsilon_S \rangle = \frac{\langle \epsilon_S \epsilon_C
\rangle}{\Omega^{1/2} [E]}.
\end{equation}
From a physical point of view, this indicates that the steady-state
concentration of substrate in the compartment is not equal to the
value predicted by the MM equation (i.e. [S]) and hence the non-zero
value of the average of the fluctuations about [S]. The real substrate
concentration inside the compartment is obtained by substituting
Eqs. (28) and (26) in Equation (14), leading to:
\begin{equation}
\left \langle \frac{n_S}{\Omega} \right \rangle = [S] + \frac{K_M
\alpha^2}{(1-\alpha)[K_M+[E_T](1-\alpha)^2]\Omega}
\end{equation}

\subsubsection*{An alternative mesoscopic rate equation replacing the MM
equation}

The renormalization of the steady-state substrate concentration
indicates the breakdown of the MM equation; this phenomenon occurs
because of non-zero correlations between noise in the substrate and
enzyme concentrations, $\langle \epsilon_S \epsilon_C \rangle$,
which the MM equation implicity neglects. To obtain the alternative
to the latter, one needs to obtain a relationship between the
normalized reaction velocity, $\alpha$ and the real substrate
concentration $\langle n_S/\Omega \rangle$; writing $[S]$ in terms
of $\alpha$ and substituting in Eq. (29), one obtains this new
relation:
\begin{align}
&\alpha + \left(1 - \frac{\langle n_S/\Omega \rangle}{K_M+\langle
n_S/\Omega \rangle}\right) f(\alpha)\Omega^{-1} =\frac{\langle
n_S/\Omega
\rangle}{K_M+\langle n_S/\Omega \rangle}, \\
&f(\alpha) = \frac{ \alpha^2}{K_M+[E_T](1-\alpha)^2}
\end{align}
Note that in the limit of large volumes, the second term on the left
hand side of Eq. (30) becomes vanishingly small and we are left
with the MM equation. In the results section the quantity on the right
hand side of Eq. (30) is referred to as $\alpha_M$ since
this is the normalized reaction velocity which would be predicted by
the MM equation given the measured substrate concentration $\langle
n_S/\Omega \rangle$ inside the compartment. A quick estimate of the
magnitude of the error that one is bound to incur by using the
conventional MM equation can be obtained by substituting $\alpha =
1/2$ (i.e. enzyme is half saturated with substrate) in Eqs.
(30) and (31), solving for $\alpha_M$ and then using this value to
compute the fractional error $e = 1 - \alpha_M/\alpha$. This leads
to the simple expression:
\begin{equation}
e = [1 + \Omega([E_T]+4K_M)]^{-1}
\end{equation}
We finish this section by noting that Eq. (30) will be found to
be valid generally and not only for the simple Michaelis-Menten
scheme treated in this section; the details of the reaction network
come in through the form of Eq. (31) which is
reaction-specific.

\subsection*{Model II: Michaelis-Menten reaction occurring in a
compartment volume of sub-micron dimensions. Substrate is input into
compartment in groups or bursts of $M$ molecules at a time.}

A natural generalization of Model I which has direct biological
application is when substrate molecules are fed into
the compartment $M$ at a time with mean rate $k_{in}^0$. The total
mean substrate input rate is then equal to $k_{in} = M k_{in}^0$.
The master equation for this process is Eq. (13) with the first
term on the right hand side replaced by $\Omega
(\Theta_S^{-M}-1)k_{in}^0$. This leads to the following change in the
expression for $a_2$ (Eq. 17):
\begin{align}
\frac{1}{2} k_{in} \frac{\partial^2}{\partial \epsilon_S^2} &
\rightarrow \frac{1}{2} k_{in} M \frac{\partial^2}{\partial
\epsilon_S^2}.
\end{align}
Note that since the expression for $a_1$ (Eq. 16) is unchanged,
the deterministic equations are precisely the same as those of Model
I. However now the fluctuations about the macroscopic substrate
concentration are enhanced by a factor $M$; consequently the entries
in the vector B in Eq. (25) need the change $k_{in}
\rightarrow k_{in} M$. The analysis proceeds in the same manner as
before. The mesoscopic rate equation replacing the MM equation is
now given by Eq. (30) together with:
\begin{equation}
f(\alpha) = \frac{ \alpha [\alpha
+\frac{1}{2}(M-1)]}{K_M+[E_T](1-\alpha)^2}.
\end{equation}
The fractional error rate evaluated at $\alpha = 1/2$ gives:
\begin{equation}
e = \frac{M}{M + \Omega([E_T]+4K_M)}
\end{equation}
This clearly shows that generally larger deviations from the
predictions of the MM equation are expected in this case compared to
those computed for Model I.

\subsection*{Model III: Michaelis-Menten reaction with competitive
inhibitor occurring in a compartment volume of sub-micron
dimensions. Substrate input as in two previous models.}

Competitive inhibition is modeled by the set of reactions:
$\xrightarrow{k_{in}} S + E \xrightleftharpoons[k_{1}]{k_0}  \ C
\xrightarrow{k_2} E + P, \ E\xrightleftharpoons[k_3]{k_4} EI$, where $k_4 = k_4^0 [I]$ and $[I]$ is the
inhibitor concentration (similar models have been studied by Roussel and collaborators \cite{Roussel1,Roussel2} in the context of biochemical oscillators though these assume $M=1$). In the rest of this section, we change the
notation of enzyme-inhibitor complex from $EI$ to $V$, just to make the
math notation easier to read. The substrate input into the compartment is considered
to occur as in Model II since this encapsulates that of Model I as
well. The master equation for this system is:
\begin{align}
\frac{d\pi}{dt} & = k_{in}^0 \Omega (\Theta_S^{-M}-1)\pi +
\frac{k_0}{\Omega} (\Theta_S \Theta_C^{-1}-1) n_S n_E \pi + k_1 (\Theta_C
\Theta_S^{-1}-1) n_C \pi \nonumber \\ &+ k_2 (\Theta_C \Theta_P^{-1} - 1) n_C \pi + k_3 (\Theta_V-1)n_V + k_4
(\Theta_V^{-1}-1)n_E.
\end{align}
The change of variables from $n_X$ to $\epsilon_X$ is done as
before, however note that now the conservation law for enzyme is different
than in the two previous models. The total enzyme concentration is now equal to
$[E_T]=[E]+[C]+[V]$ from which it follows that $n_E = \Omega [E]
-\Omega^{1/2}(\epsilon_C+\epsilon_V)$. The description is chosen to
be in terms of numbers of molecules of types $C$, $S$ and $V$ and
thus $E$ being a dependent variable does not show up explicitly in
the step operators of the master equation above.

Due to the significant number of changes in the terms of the expansion
from those of previous models, we will show the equivalent of
Eqs. (15)-(18) in full. The master equation in the new variables
$\epsilon_X$ is given by:

\begin{align}
\frac{\partial \Pi}{\partial t} - \Omega^{1/2}
\biggl(\frac{d[C]}{dt} \frac{\partial \Pi}{\partial \epsilon_C} +
\frac{d[P]}{dt} \frac{\partial \Pi}{\partial \epsilon_P} +
\frac{d[S]}{dt} \frac{\partial \Pi}{\partial \epsilon_S} &+
\frac{d[V]}{dt} \frac{\partial \Pi}{\partial \epsilon_V} \biggr) = \nonumber \\
&\Omega^{1/2} a_1 \Pi + \Omega^{0} a_2 \Pi + \Omega^{-1/2} a_3 \Pi +
O(\Omega^{-1})
\end{align}
where
\begin{align}
a_1 = &-(k_{in} + k_1 [C] - k_0 [E][S]) \frac{\partial}{\partial
\epsilon_S} + ((k_1+k_2)[C] - k_0 [E][S])\frac{\partial }{\partial
\epsilon_C} \\ \nonumber &+ (k_3 [V] - k_4 [E])
\frac{\partial}{\partial \epsilon_V} - k_2[C] \frac{\partial}{\partial \epsilon_P},
\end{align}

\begin{align}
\nonumber a_2 = & \frac{1}{2}k_{in} M
\frac{\partial^2}{\partial\epsilon_S^2} + \frac{1}{2}
\biggl(\frac{\partial}{\partial \epsilon_S}-\frac{\partial}{\partial
\epsilon_C}\biggr)^2 (k_0 [S][E]+k_1 [C]) + k_2 \biggl[\frac{\partial}{\partial \epsilon_C} - \frac{\partial}{\partial \epsilon_P} \biggr] \epsilon_C \nonumber \\ &+
\biggl[\frac{\partial}{\partial \epsilon_S} -
\frac{\partial}{\partial \epsilon_C} \biggr] [k_0(\epsilon_S
[E]-(\epsilon_C+\epsilon_V) [S])- k_1 \epsilon_C] + k_3 \biggl(
\frac{\partial}{\partial \epsilon_V} \epsilon_V + \frac{1}{2}
\frac{\partial^2}{\partial \epsilon_V^2} [V] \biggr) \nonumber \\ &+
k_4 \left( \frac{1}{2} [E] \frac{\partial^2}{\partial \epsilon_V^2}
+ \frac{\partial}{\partial\epsilon_V} (\epsilon_C + \epsilon_V)
\right) + \frac{1}{2} k_2 \biggl(\frac{\partial}{\partial \epsilon_P}-\frac{\partial}{\partial \epsilon_C}\biggr)^2 [C],
\end{align}

\begin{align}
\nonumber a_3 = &\frac{1}{2} \biggl(\frac{\partial}{\partial
\epsilon_S}-\frac{\partial}{\partial \epsilon_C}\biggr)^2(k_0
\epsilon_S [E] - k_0 (\epsilon_C + \epsilon_V) [S] + k_1 \epsilon_C)
- k_0 \biggl[\frac{\partial}{\partial \epsilon_S}
-\frac{\partial}{\partial \epsilon_C} \biggr] \epsilon_S (\epsilon_C
+ \epsilon_V) \nonumber \\&+ \frac{1}{2} k_2 \biggl(\frac{\partial}{\partial
\epsilon_P}-\frac{\partial}{\partial \epsilon_C}\biggr)^2 \epsilon_C + \frac{1}{2} k_3
\frac{\partial^2}{\partial \epsilon_V^2} \epsilon_V  - \frac{1}{2}
k_4 \frac{\partial^2}{\partial\epsilon_V^2} (\epsilon_C + \epsilon_V).
\end{align}

\subsubsection*{Analysis of $\Omega^{1/2}$ terms}

As for previous models, these terms give the macroscopic equations.
Equating both terms of this order on the right and left hand sides
of Eq. (37) and using Eq. (38), one obtains:
\begin{align}
{d[S]}/{dt} & = k_{in} - k_0 [E][S] + k_1 [C], \\
{d[C]}/{dt} & = k_0 [E][S] - (k_1+k_2)[C], \\
{d[P]}/{dt} & = k_2 [C], \\
{d[V]}/{dt} & = k_4 [E] - k_3 [V].
\end{align}
In the steady-state we have the Michaelis-Menten (MM) equation:
\begin{equation}
\frac{d[P]}{dt} = \frac{v_{max}[S]}{K_M(1+\beta)+[S]},
\end{equation}
where $\beta = [I]/K_i$ and $K_i=k_3/k_4^0$ is the dissociation
constant of the inhibitor.

\subsubsection*{Analysis of $\Omega^{0}$ and $\Omega^{-1/2}$ terms}

The equations for the first moments are easily obtained and we shall
not reproduce them here; suffice to say that at steady-state, it is
found that $\langle \epsilon_{S,C,V} \rangle \rightarrow 0$ which
implies that to this order in the system-size expansion there cannot
be any corrections to the macroscopic equations or to the MM
equation. The addition of a new species, $V$, does however
substantially increase the algebraic complexity in the equations of
motion for the second moments computed using terms up to order $\Omega^{0}$.
In particular the matrix A is now a 6 $\times$ 6 matrix, rather
than the 3 $\times$ 3 matrix of the previous two models.

\begin{equation}
\partial_t \left[\begin{array}{c} \langle \epsilon_S^2 \rangle \\ \langle
\epsilon_C^2 \rangle \\ \langle \epsilon_V^2 \rangle \\ \langle
\epsilon_{S}
\epsilon_{V} \rangle \\ \langle \epsilon_{C} \epsilon_{V} \rangle \\
\langle \epsilon_{S} \epsilon_{C} \rangle
\end{array}\right]=A \cdot \left[\begin{array}{c} \langle \epsilon_S^2 \rangle \\ \langle
\epsilon_C^2 \rangle \\ \langle \epsilon_V^2 \rangle \\ \langle
\epsilon_{S}
\epsilon_{V} \rangle \\ \langle \epsilon_{C} \epsilon_{V} \rangle \\
\langle \epsilon_{S} \epsilon_{C} \rangle
\end{array}\right] + B,
\end{equation}
where,
\begin{equation}
A = \left(\begin{array}{cccccc} -2k_0 [E] & 0 & 0 & 2k_0 [S] & 0 & 2(k_1 + k_0 [S])\\
0 & -2k_0(K_M+ [S]) & 0 & 0 & -2k_0 [S] & 2k_0 [E] \\
0 & 0 & -2k' & 0 & -2k_4 & 0 \\
0 & -k_4 & 0 & -k' & -(k_2+k') & -k_4 \\
0 & -k_4 & -k_0[S] & k_0 [E] & -k_0(K_M+[S])-k' & 0 \\
0 & -k_2 & 0 & 0 & 0 & -k_2
\end{array}\right),
\end{equation}
and
\begin{equation}
B = \left(\begin{array}{c} k_{in}M+k_1[C]+k_0[S][E] \\
k_0 ([S][E] + K_M[C]) \\
k_4 [E] + k_3 [V] \\
0 \\
0 \\
\frac{1}{2}(k_{in}M + k_2 [C]) \end{array}\right).
\end{equation}
In the above equations we have defined $k'=k_3+k_4$. Note also that
the system of equations has been simplified through the application
of a few row operations.

Now to next order, i.e. $\Omega^{-1/2}$, the first moments of the
concentrations of molecules of type $C$ and $V$ are governed by the
equation of motions:
\begin{align}
{d \langle \epsilon_C \rangle}/{dt}  &= -k_0([S] + K_M) \langle
\epsilon_C\rangle - k_0[S]\langle \epsilon_V\rangle + k_0 [E]
\langle \epsilon_S \rangle - k_0 \Omega^{-1/2} (\langle \epsilon_S
\epsilon_C \rangle + \langle \epsilon_S \epsilon_V \rangle) \\
{d \langle \epsilon_V \rangle}/{dt} &= -k_3 \langle \epsilon_V
\rangle - k_4 (\langle \epsilon_C \rangle + \langle \epsilon_V
\rangle)
\end{align}
As in previous models, since the production of product \emph{P} from
complex occurs through a decay process, it follows that at
steady-state, $\langle \epsilon_C \rangle = 0$ which also implies
$\langle \epsilon_V \rangle = 0$ from Eq. (50). Hence it
follows from Eq. (49) that $\langle \epsilon_S \rangle =
[\langle \epsilon_S \epsilon_C \rangle + \langle
\epsilon_S \epsilon_V \rangle]/\Omega^{1/2}[E]$. The two cross
correlators can be estimated to order $\Omega^0$ by solving
Eqs. (46)-(48). The non-zero value of $\langle \epsilon_S \rangle$ implies
a renormalization of the substrate concentration inside
the compartment and hence to a new rate equation replacing the MM equation. This
is obtained exactly in the same manner as previously shown for Model I. The mesoscopic rate
equation is found to be given by Eq. (30) together with:
\begin{equation}
f(\alpha) = \frac{(1+\beta)}{K_M [E_T]} \frac{\sum_{i=0}^4 c_i
(1-\alpha)^i}{\sum_{i=0}^4 d_i (1-\alpha)^i},
\end{equation}
where the numerator coefficients are given by:
\begin{align}
c_0 = &+k_3(\beta+1)^3K_M^2 k_0 [E_T], \\
c_1 = &+K_M (\beta+1)^2[(\beta+1) [E_T] k_3^2-(3 \beta+2) [E_T]
k_0 K_M k_3+k_0 \beta v_{max} K_M], \\
c_2 = &-K_M (\beta+1) [2(2 \beta+\beta^2+1) [E_T] k_3^2-((3
\beta^2+4\beta+1) [E_T] k_0 K_M+\\ \nonumber &-\beta(\beta+1)
v_{max}-k_0 [E_T]^2) k_3+\beta(1+2 \beta) k_0 v_{max}
K_M-(\beta+1)[E_T] k_0 v_{max}], \\
c_3 = &+[(1+3\beta+\beta^3+3\beta^2)[E_T] K_M
k_3^2-(\beta(\beta+1)^2[E_T] k_0 K_M^2+
\\ \nonumber &(-\beta(\beta+1)^2 v_{max}-2(1+\beta)[E_T]^2
k_0) K_M+\beta(\beta+1)[E_T] v_{max}) k_3+ \\
\nonumber &\beta^2(1+\beta)k_0 v_{max}
K_M^2-(2+3\beta+2\beta^2)[E_T] k_0 v_{max} K_M], \\
c_4 = &-[(-(\beta+1)[E_T]^2 k_0
K_M+\beta(\beta+1)[E_T] v_{max}) k_3+ \\
\nonumber &-[E_T](\beta+\beta^2+1)k_0 v_{max} K_M],
\end{align}
and the denominator coefficients by:
\begin{align}
d_0 = &+K_M^2 k_0 k_3 (1+\beta)^4 , \\
d_1 = &+K_M k_3 (\beta+1)^3 [\beta (k_3-k_0 K_M)+k_3],\\
d_2 = &+K_M k_0 (\beta+1)^2 [k_3 [E_T](\beta
+2)+v_{max}], \\
d_3 = &+(\beta+1) [k_3^2 \beta^2 [E_T]-k_0\beta^2 k_3 K_M [E_T]+2
k_3^2 \beta [E_T]\\&-k_0 k_3 \beta K_M [E_T]-k_0
\beta v_{max} K_M+k_3^2 [E_T]], \\
d_4 = &+ [E_T] k_0 [k_3 [E_T]+k_3 \beta [E_T]+v_{max}].
\end{align}

Note that $\sum_{i=0}^4 c_i =0$ such that at $\alpha = 0$, there
is no correction to the MM equation i.e. $\alpha_M = 0$ also. The
case $\beta=0$ reduces to Model II, i.e. $f(\alpha)$ is given by
Eq. (34).

\subsection*{Stochastic simulation}

In this section we briefly describe the simulation methods used to
verify the theoretical results which are described in detail in the
Results section. All simulations were carried out using Gillespie's exact
stochastic simulation algorithm, conveniently implemented in the
standard simulation platform, Dizzy \cite{SSA}.

The data points in Figure 2 were generated by iterating the following four-step procedure: (i) pick a value for $\alpha$ between 0 and 1. This gives the substrate input rate $k_{in} = \alpha v_{max}$; (ii) run the
simulation and measure the ensemble-averaged substrate concentration, $\langle n_s /\Omega \rangle = [S^*]$ at long
times; (iii) compute $\alpha_M$ using the MM equation,
$\alpha_M=[S^*]/([S^*]+K_M)$; (iv) compute the absolute percentage error $R_p =
100 |(1-\alpha_M/\alpha)|$. The solid curves in Figure 2 were obtained by numerically solving the cubic
polynomial in $\alpha$ given by Eqs. (7) and (8) in the Results section for given values of $\alpha_M$ and
then using the above expression for $R_p$. Figure 3 is generated in the same manner as Figure 2, except that: in step (i) we fix $M$ and pick a value for $\alpha$ between 0 and 1. Since $k_{in} = M k_{in}^0$, the required simulation parameter is $k_{in}^0 = \alpha v_{max} / M$; step (iv) is not needed. The solid curves were obtained by numerically solving the cubic polynomial in $\alpha$ given by Eqs. (7) and (9) in the Results section for given values of $[S^*]$.
The y-axis for this figure is $v/v_{max} = \alpha_M$ for the MM equation and $v/v_{max} = \alpha$ for the stochastic model. Figure 4 is obtained by numerically solving the quintic polynomial in $\alpha$ given by Eqs. (7) and (12) in the Results section together with the coefficients given by Eqs. (52)-(62) in the present section; the inhibitor concentration, $[I]$, is varied while the substrate concentration, $[S^*]$, is kept fixed. The substrate concentration is chosen so that at $[I] = 0$, $v/v_{max}=0.909$ in all cases. Note that for models I and II, $\alpha_M=[S^*]/([S^*]+K_M)$ while for Model III, $\alpha_M=[S^*]/([S^*]+(1+\beta) K_M)$. Note that the error bars are very small on the scale of the figures and thus are not shown.

\section*{Acknowledgments} It is a pleasure to thank Arthur Straube and Philipp Thomas for interesting discussions. The author gratefully acknowledges support from SULSA (Scottish Universities Life Sciences Alliance).

\newpage

\begin{figure}
\includegraphics [width=6in]{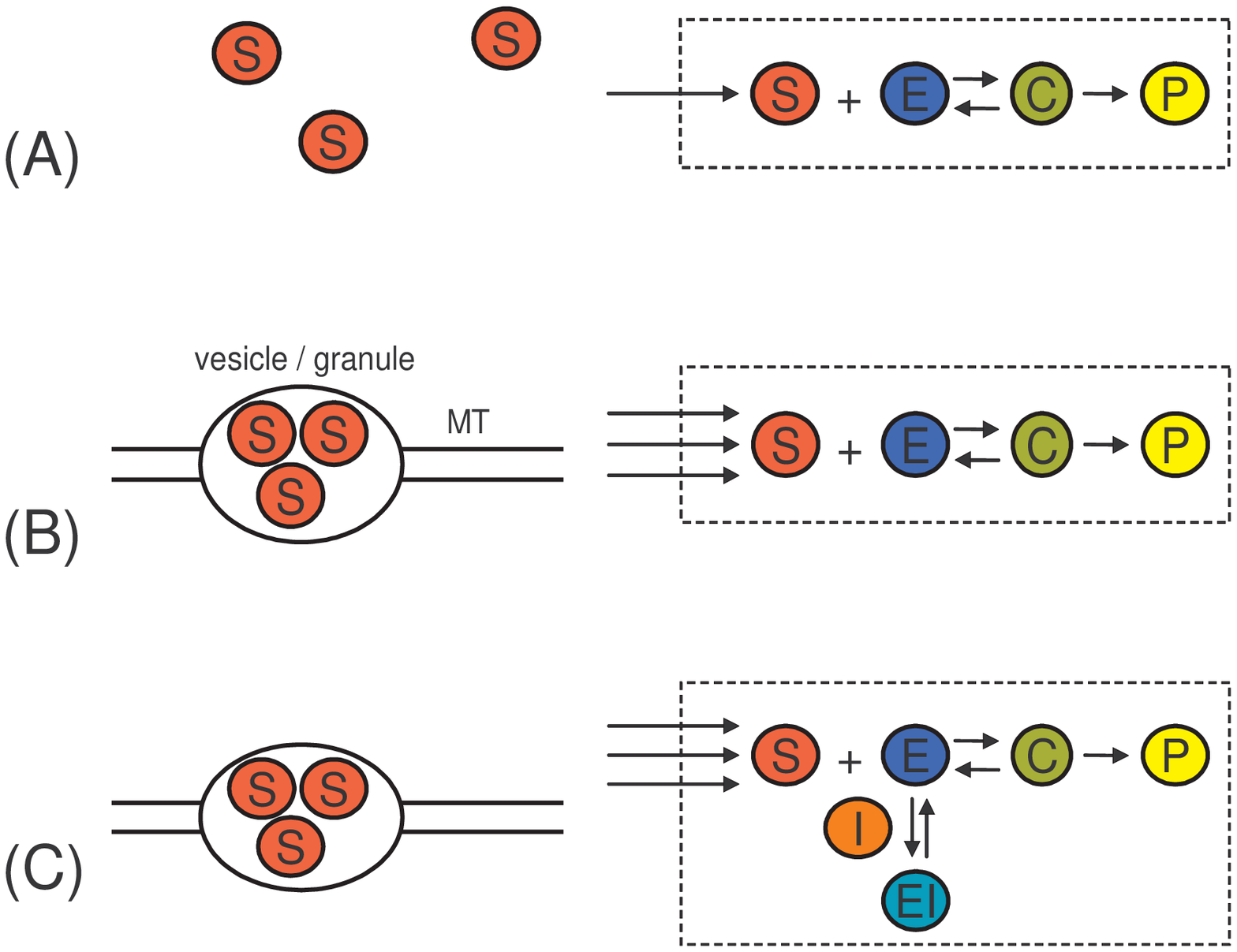}
\caption{\textbf{Schematic illustrating the three models considered in this article.}
(A) Model I: Michaelis-Menten reaction occurring in a compartment volume of sub-micron dimensions (shown by dashed rectangle). Substrate input into compartment occurs via a Poisson process i.e. diffusion-mediated substrate transport. (B) Model II: As for Model I but now substrate is input into compartment in groups or bursts of \emph{M} molecules at a time i.e. vesicle-mediated substrate transport along microtubules (MT). (C) Model III: Michaelis-Menten reaction with competitive inhibitor (\emph{I}) occurring in a small subcellular compartment. Substrate transport as in previous two models.}
\end{figure}

\begin{figure}
\includegraphics [width=5in]{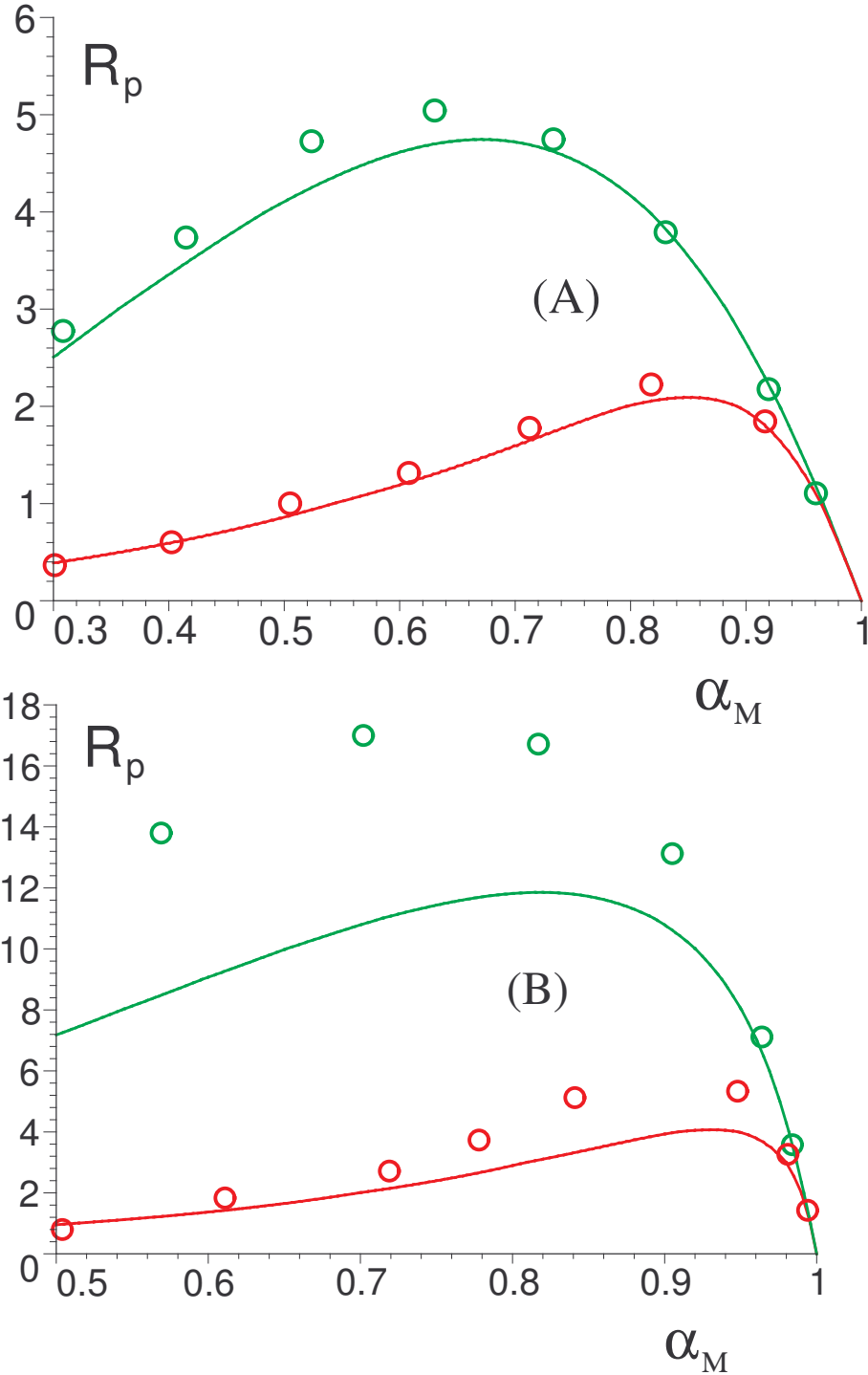}
\caption{\textbf{Deviations from the predictions of the MM equation for diffusion-mediated substrate transport.}
(Model I) Plot of the Percentage Error in reaction velocity,
$R_p = 100|1 - \alpha_M/\alpha|$, versus the normalized reaction
velocity of the MM equation, $\alpha_M$ for 10 enzymes (green) and
100 enzymes (red) with $K_M = 10\mu M$ in compartments with diameter
$100$nm (A) and $50$nm (B) . The solid lines show the theoretical
predictions, as encapsulated by Eqs. (7) and (8); the data points
are obtained by stochastic simulation (see Methods for details). }
\end{figure}

\begin{figure}
\includegraphics [width=6in,]{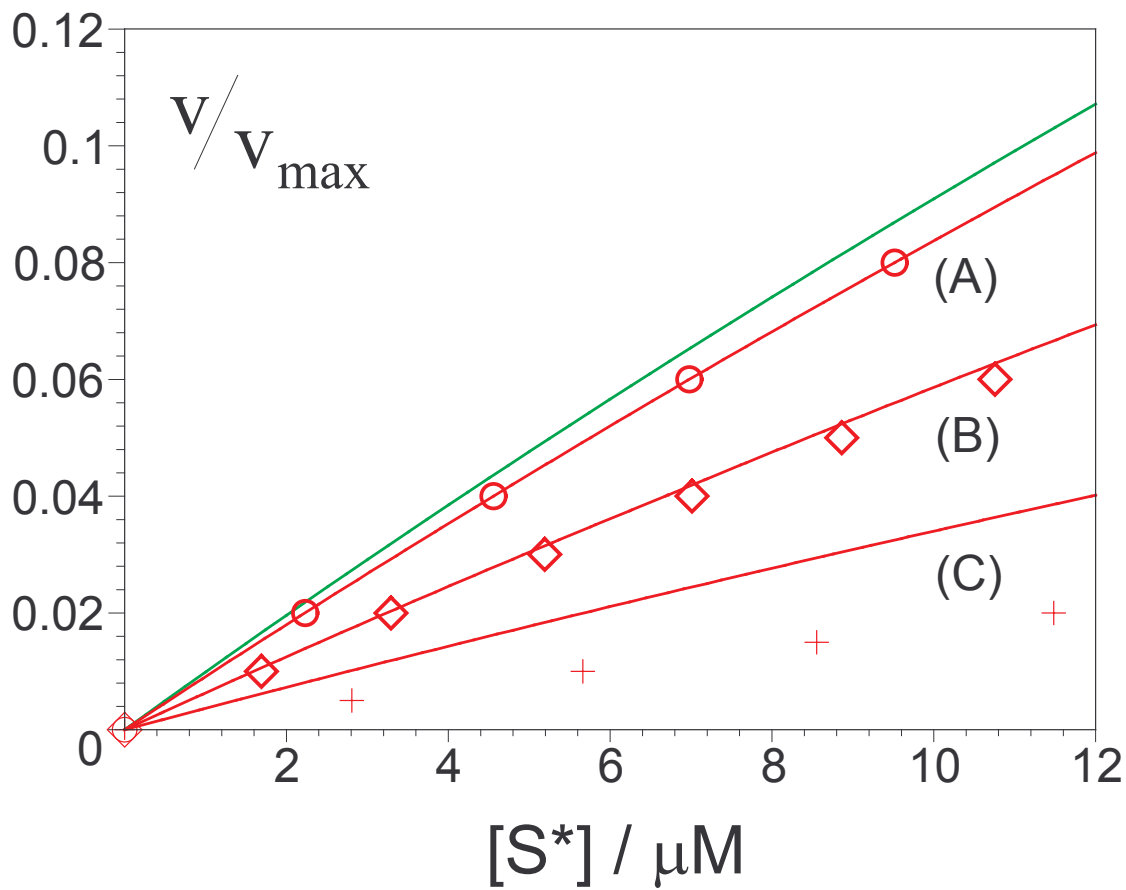}
\hfill \caption{\textbf{Deviations from the predictions of the MM equation for vesicle-mediated substrate transport.}
(Model II) Testing the validity of the MM relationship at small substrate concentrations for the case in which
substrate input into compartments occurs in bursts. The data is for
10 enzymes with $K_M = 100 \mu M$ in compartments of diameter (A)
200nm (circles), (B) 100nm (diamonds) and (C) 50nm (crosses);
substrate is input $M = 50$ molecules at a time. The deterministic
prediction for all three cases is the same MM equation shown by
the green curve. In contrast, the stochastic models, [Eqs. (7)
and(9)], predict different rate equations for each case (red solid lines).
Data points are obtained by stochastic simulation (see Methods for details).
Note that $v/v_{max} = \alpha_M$ and $\alpha$ for solid green and red lines respectively.}
\end{figure}

\begin{figure}
\includegraphics [width=6in,]{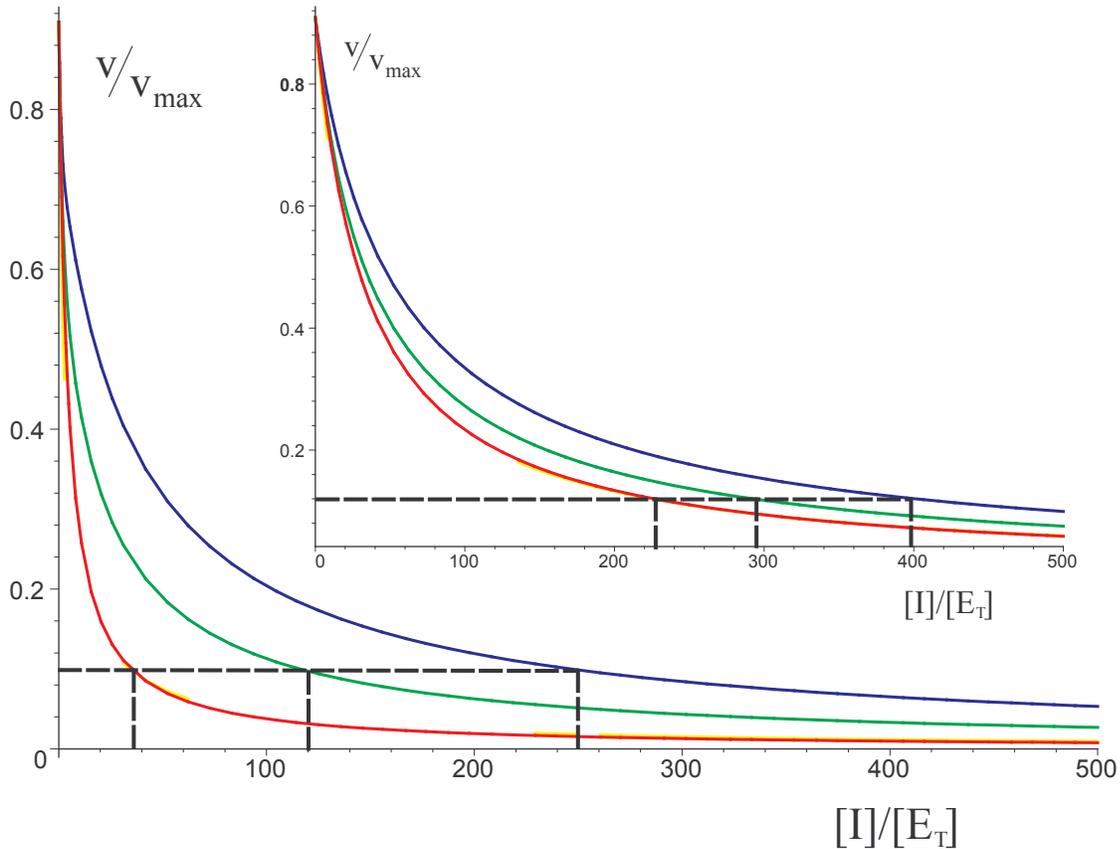}
\hfill \caption{\textbf{Effects of intrinsic noise on the inhibition of enzyme activity in small compartments.}
(Model III) Plots of normalized enzyme activity
versus normalized inhibitor concentration (measured in units of the total enzyme
concentration $[E_T]$) for 10 enzymes with $K_M = 100 \mu M$
in compartments of diameter 50nm and 100nm (inset). The colors
correspond to: (red) MM equation; (green) stochastic model, $M =
20$; (blue) stochastic model, $M = 50$. The latter two curves are those predicted
by theory [Eqs. (7) and(12)]. Parameters same as mentioned in
caption of Table 3 (except for [I], which is a variable in the
present case). Substrate concentrations chosen so that at $[I] = 0$,
$v/v_{max}=0.909$ in all cases. Black dashed lines contrast the inhibitor
concentration required to decrease enzyme activity from 0.909 to 0.1 as
predicted by the MM equation and the stochastic models. Note that
$v/v_{max} = \alpha_M$ and $\alpha$ for solid red and blue/green lines respectively.}
\end{figure}

\newpage

\begin{table}[h]
\caption{Maximum Percentage error in reaction velocity from prediction of the MM equation for Model I. The copy number indicates the total number of enzyme molecules per compartment. Values in bold and in square brackets are those estimated by simulation; the italic values are obtained from the derived theoretical expressions, Eqs. (7) and (8).}
\centering 

\begin{tabular}{c c c c c} 
\\
\hline \hline 
\\
D/nm & $K_M=10\mu M$ & $100\mu M$ & $1000\mu M$ & Copy No. \\ [0.5ex]
\\
50 & \emph{11.83} \textbf{[17.00]} & \emph{4.09} \textbf{[4.33]} & \emph{0.59} & 10 \\ 
100 & \emph{4.74} \textbf{[5.00]} & \emph{0.73} \textbf{[0.74]} & \emph{0.08} & 10\\
200 & \emph{0.90} & \emph{0.10} & \emph{0.01} & 10\\ [1ex] 
\hline \hline 
\\
50 & \emph{3.98} \textbf{[5.33]} & \emph{1.88} \textbf{[2.02]} & \emph{0.43} & 100 \\ 
100 & \emph{2.10} \textbf{[2.23]} & \emph{0.52 }\textbf{[0.52]} & \emph{0.07} & 100\\
200 & \emph{0.61} & \emph{0.09} & \emph{0.01} & 100\\ [1ex] 
\hline 
\end{tabular}
\end{table}

\begin{table}[h]
\caption{Maximum Percentage error in reaction velocity from prediction of the MM equation for Model II. The copy number indicates the total number of enzyme molecules per compartment. Values in bold and in square brackets are those estimated by simulation; the italic values are obtained from the theoretical expressions, Eqs. (7) and (9).}
\centering 
\begin{tabular}{c c c c c} 
\\
\hline \hline 
\\
D/nm & $K_M=10\mu M$ & $100\mu M$ & $1000\mu M$ & Copy No. \\ [0.5ex]
\\
50 & \emph{225.40} & \emph{152.83} \textbf{[291.56]} & \emph{45.43} &  10\\ 
100 & \emph{161.59} \textbf{[331.66]} & \emph{52.74} \textbf{[58.39]} & \emph{6.82} \textbf{[6.99]} & 10\\
200 & \emph{65.09} & \emph{8.45} \textbf{[8.50]} & \emph{0.88} & 10\\ [1ex] 
\hline \hline 
\\
50 & \emph{32.97} & \emph{30.17} \textbf{[61.66]} & \emph{18.14} & 100 \\ 
100 & \emph{30.78} \textbf{[66.03]} & \emph{19.76 }\textbf{[24.52]} & \emph{5.57} \textbf{[6.06]} & 100\\
200 & \emph{21.27} & \emph{6.61} \textbf{[6.91]} & \emph{0.85} & 100\\ [1ex] 
\hline 
\end{tabular}
\end{table}

\begin{table} [h]
\caption{Maximum Percentage error in reaction velocity from prediction of the MM equation for Model III.
The total number of enzyme molecules per compartment is ten in all cases. Values in bold and in square brackets
are those estimated by simulation; the italic values are obtained from the theoretical expressions, Eqs. (7) and (12).
The parameters are: $k_0=10^9 s^{-1} M^{-1}, k_1 = k_3 = 1000 s^{-1}, k_4^0 = 10^7 s^{-1} M^{-1}$, and $[I]=10[E_T]$.} 
\centering 
\begin{tabular}{c c c c c} 
\\
\hline \hline 
\\
D/nm & $K_M=10\mu M$ & $100\mu M$ & $1000\mu M$ & M (burst size) \\ [0.5ex]
\\
50 & \emph{67.8} \textbf{[76.8]} & \emph{67.8} \textbf{[76.5]} & \emph{67.8} &  1\\ 
100 & \emph{20.8} \textbf{[26.4]} & \emph{20.6} \textbf{[26.1]} & \emph{20.6} & 1\\
200 & \emph{2.8} & \emph{2.7} & \emph{2.7} & 1\\ [1ex] 
\hline \hline 
\\
50 & \emph{1001.8} & \emph{234.9} \textbf{[169.4]} & \emph{86} & 50 \\ 
100 & \emph{343.7} \textbf{[345.5]} & \emph{73.4}\textbf{[75.2]} & \emph{26.2} \textbf{[31.5]} & 50\\
200 & \emph{71.4} & \emph{11.3} \textbf{[11.5]} & \emph{3.6} & 50\\ [1ex] 
\hline 
\end{tabular}
\end{table}


\begin{thebibliography} {20}

\bibitem{LubyPhelps1} Luby-Phelps K: \textbf{Cytoarchitecture and Physical properties of Cytoplasm: Volume, Viscosity, Diffusion, Intracellular Surface Area.} {\it Int Rev Cytology} 1999, \textbf{192}:189--221
\bibitem{Alberts} Alberts B et al: Molecular Biology of the Cell. Garland Publishing; 1994
\bibitem{Minton} Minton AP: \textbf{How can biochemical reactions within cells differ from those in test
tubes?} {\it J Cell Sci} 2006, \textbf{119}:2863–-2869
\bibitem{Trepat} Trepat X et al: \textbf{Universal physical responses to stretch in the living cell.} {\it Nature} 2007, \textbf{447}:592–-596
\bibitem{SchnellTurner} Schnell S, Turner TE: \textbf{Reaction kinetics in intracellular environments with macromolecular crowding: simulations and rate laws.} {\it Prog Biophys Mol Biol} 2004, \textbf{85}:235--260
\bibitem{Medalia} Medalia O et al: \textbf{Macromolecular Architecture in Eukaryotic Cells Visualized by Cryoelectron tomography.} {\it Science} 2004, \textbf{298}:1209--1213
\bibitem{LubyPhelps2} Luby-Phelps K, Castle PE, Taylor DL, Lanni F: \textbf{Hindered diffusion of inert tracer particles in the cytoplam of mouse 3T3 cells.} {\it Proc Natl Acad Sci USA} 1987, \textbf{84}:4910--4913
\bibitem{Pick} Pick H et al: \textbf{Investigating Cellular Signaling Reactions in Single Attoliter vesicles.} {\it J Am Chem Soc} 2005, \textbf{127}:2908--2912
\bibitem{Grima1} Grima R, Schnell S: \textbf{Modelling reaction kinetics inside cells.} {\it Essays in Biochemistry} 2008, \textbf{45}:41--56
\bibitem{Gillespie} Gillespie DT: \textbf{Exact stochastic simulation of coupled chemical reactions}, {\it J Phys Chem} 1977, \textbf{81}:2340--2361
\bibitem{AvrahamHavlin} Ben-Avraham D, Havlin S: {\it Diffusion and Reactions in Fractals and Disordered Systems}. Cambridge University Press; 2000
\bibitem{VanKampen} van Kampen NG: {\it Stochastic processes in physics and chemistry}. Amsterdam: Elsevier; 2007
\bibitem{Bartholomay1} Bartholomay AF: \textbf{A Stochastic Approach to Statistical Kinetics with Application to Enzyme Kinetics}, {\it Biochemistry} 1962, \textbf{1}:223--230
\bibitem{Bartholomay2} Bartholomay AF: \textbf{Enzymatic Reaction-Rate Theory - A Stochastic Approach}, {\it Annals of the New York Academy of Sciences} 1962, \textbf{96}:897
\bibitem{Jachimowski} Jachimowski CJ, McQuarrie DA, Russell ME: \textbf{A Stochastic Approach to Enzyme-Substrate Reactions}, {\it Biochemistry} 1964, \textbf{3}:1732--1736
\bibitem{Grima2} Grima R: \textbf{Multiscale modeling of biological pattern formation} {\it Curr Top Dev Biol} 2008, \textbf{81}:435--460
\bibitem{Berg} Berg JM, Tymoczko JL, Stryer L {\it Biochemistry} $5^{th}$ edition. New York: Freeman; 2002
\bibitem{Popov} Popov S, Poo MM: \textbf{Diffusional transport of macromolecules in developing nerve processes.} {\it J Neurosci} 1992, \textbf{12}:77--85
\bibitem{ArrioDupont} Arrio-Dupont M, Cribier S, Foucault G, Devaux PF, d'Albis A: \textbf{Diffusion of fluorescently labelled macromolecules in cultured muscle cells.} {\it Biophys J} 1996, \textbf{70}:2327--2332
\bibitem{Ainger} Ainger K et al: \textbf{Transport and localization of exogenous myelin basic protein mRNA microinjected into Oligodendrocytes.} {\it J Cell Biol} 1993, \textbf{123}:431–-441
\bibitem{BassellSinger} Bassell G, Singer RH: \textbf{mRNA and cytoskeletal filaments.} {\it Curr Opin Cell Biol} 1997, \textbf{9}:109--115
\bibitem{Fersht} Fersht A {\it Structure and mechanism in protein science}. New York: Freeman; 1998
\bibitem{Baras} Baras F, Mansour MM: \textbf{Reaction-diffusion master equation: A comparison with microscopic simulations.} {\it Phys Rev E} 1996, \textbf{54}:6139--6148
\bibitem{English} English BP, Min W, van Oijen AM, Lee KT, Luo G et al:\textbf{Ever-fluctuating single enzyme molecules: Michaelis-Menten equation revisited.} {\it Nat Chem Biol} 2005, \textbf{2}:87--94
\bibitem{Kou} Kou SC, Cherayil BJ, Min W, English BP, Xie SX: \textbf{Single-molecule Michaelis-Menten
equations.} {\it J. Phys. Chem B} 2005, \textbf{109}:19068--19081
\bibitem{Stefanini} Stefanini MO, McKane AJ, Newman TJ: \textbf{Single enzyme pathways and substrate fluctuations.} {\it Nonlinearity} 2005, \textbf{18}:1575--1595
\bibitem{Qian} Qian H, Elson EL: \textbf{Single-molecule enzymology: stochastic Michaelis-Menten kinetics.} {\it Biophys Chem} 2002, \textbf{101-102}:565--576
\bibitem{Grima3} Grima R: \textbf{Noise-induced breakdown of the Michaelis-Menten equation in steady-state conditions.} {\it Phys Rev Letts} 2009, \textbf{102}:218103
\bibitem{McKane} McKane AJ, Nagy J, Newman TJ, Stefanini M: \textbf{Amplified biochemical oscillations in cellular systems} {\it J Stat Phys} 2007, \textbf{128}:165--191
\bibitem{Roussel1} Ngo LG, Roussel MR: \textbf{A new class of biochemical oscillators based on competitive binding} {\it Eur J Biochem} 1997, \textbf{245}:182--190
\bibitem{Roussel2} Davis KL, Roussel MR: \textbf{Optimal observability of sustained stochastic competitive inhibition oscillations at organellar volumes} {\it FEBS journal} 2006, \textbf{273}:84--95
\bibitem{SSA} Ramsey S, Orrell D, Bolouri H: \textbf{Dizzy: Stochastic simulation of large-scale genetic regulatory networks} {\it J Bioinform Comput Biol} 2005, \textbf{3}: 415--436
\end{thebibliography}
\end{document}